\documentstyle[12pt,epsf]{article}
\newcommand{\bc}{\begin{center}}
\newcommand{\ec}{\end{center}}
\newcommand{\bd}{\begin{displaymath}}
\newcommand{\ed}{\end{displaymath}}
\newcommand{\be}{\begin{equation}}
\newcommand{\ee}{\end{equation}}
\newcommand{\ba}{\begin{array}}
\newcommand{\ea}{\end{array}}
\newcommand{\bea}{\begin{eqnarray}}
\newcommand{\eea}{\end{eqnarray}}
\newcommand{\bt}{\begin{tabular}}
\newcommand{\et}{\end{tabular}}

\newcommand{\bp}{\begin{picture}}
\newcommand{\ep}{\end{picture}}
\newcommand{\bfi}{\begin{figure}}
\newcommand{\efi}{\end{figure}}

\begin{document}

\title{\huge \bf {Imaginary part of action, Future functioning as hidden 
variables. }}

\author{ 
H.B.~Nielsen ${}^{1}$ \footnote{\large\, hbech@nbi.dk} \\[5mm]
\itshape{${}^{1}$ The Niels Bohr Institute, Copenhagen, Denmark}}

\date{}

\maketitle

\begin{abstract}
A model - by myself and Masao Ninomiya -, which in principle predicts 
the initial conditions 
in a way as to minimze a certain functional of the history of the 
Universe through both past and future - a functional conceived of as 
an imaginary part of the action -  is suggested to be also helpfull 
in solving some problems for quantum mechanics. Especially as 
our model almost makes it possible in principle to calculate the full history
of the universe, it even makes it in principle calculable, which one among 
several measurement results in a quantum experiment will actually 
be realized!

Our ``complex action model'' thus is a special case of superdeterminism
- in Bells way - and does not have true causality, but rather even in 
some cases true backward causation. In fact we claim in our model that 
the SSC(Superconducting Supercollider) were stopped by the US Congress
due to the backward causation from the big amounts of Higgs particles, which
 it 
would have produced, if it had been allowed to run. The noumenon (``das Ding
an sich'') in our model is the Feyman path integrand or better 
some fundamental
quantities determined from second order effects of the latter integrand.   
\end{abstract}

% This is the beginning work on the proceedings for the Vexjo 
% conference:

\section{Introduction}

Since about 1974 I and my collaborators have worked on the very 
ambitious project or dream of Random Dynamics\cite{RDfirst}\cite{RD}
\cite{RDrev}\cite{Origin}\cite{Bled}, which intends to 
realize the hope of argueing, that almost whatever very complicated theory 
were true, it should almost certainly (in the mathematical sense 
of modulo a probabibility nul-set) lead to the laws of nature as we 
know them. That is to say we want to show that an essentially random 
complicated world-machinery would lead to mainly the Standard Model.

In realizing this project in the logically first steps we have in mind 
to think of the world machinery as some very complicated - but still with 
eneourmously many regularities - ``mathematical structure''. Then we 
propose to make some general considerations about, how one describes some 
of the features of such a structure by defining - or at least think of -
``substructures'' which could then be counted in principle, so that the 
numbers 
of the variuos possible substructures could give rise to a description.
This description would now be expressed in a more general language, i.e. not 
so 
specific to the (by God) given complicated mathematical structure,
which makes up the world machinery or the world, but rather in the terms 
of numbers as just suggested. Note how we here  made an attempt
to argue that very very generally we could tild our description so as to get 
the language of the original ``complicated mathematical structure with 
a huge amount of regularity'' replaced by some counting terminology, which 
we have ourself introduced from the human culture of mathematics. If we 
continue steps of this sort we might hope to have at the end almost talked 
the original model (Gods model) out and got it replaced by our own 
- from mathematical culture say - reformulation so that at the end we
may get a result essentially no more dependent on the original input 
``mathematical strucuture'' but rather on how we have chosen to describe it 
approximately. You should have in mind that in this way we would say we have 
hope that the end is (almost at least) independent of the original 
God-given model!    

Since we already know phenomenologically, that the world we live in is 
described by quantum mechanics causing the wellknown troubles 
in finding a good ontological basis for the world, it seems to be a bad
proposal to make a world macinery model in which the structure is in 
a certain way, like our ``mathematical structure''.(So it would 
suggestable not use such an ontological picture but rather say take 
symmetries to be 
the foundation \cite{BohrUlfbeck}) We would namely 
in quantum mechanics rather think of the wave function - which is 
what we might at  first think of as an ontological basis - as 
a mathematical object describing in a mixed way {\em both} the world 
and what we know about it, rather than just the genuinely existing world.
One would therefore a priori think that in order that our proposal 
of imagining an ontologically truly existing - complicated or not - mathematial
structure to be what the world is made from, should have a chance to match 
with 
quantum 
mechanics we would have to somehow modify it so as to say that it also 
has some information not really about the world proper, but about what we 
know about it. But with such an interpretaion of a combined information 
about ``das Ding an sich'' and our state of knowledge would not really 
be a 100\% fundamental world machinery. We shall in the present article 
consider such an interpretation of the information on the ``Ding an sich''
and the knowledge of ours being combined in a non-seperable way 
as not being so estetically attractive,  and thus we want 
to truly think of our ``complicated mathematical structure'' as being 
{\em only} ``das Ding an sich''. But hen it seems that we have almost brought 
our model to being killed immediately by making itself open essentially 
to a nogo-theorem, saying that such a type of model cannot match with 
quantum mechanics.

One might say that the EPR\cite{EPR} argumentation is precisely that 
at least in a local 
theory 
with such existing features cannot exist.

The way we shall solve this problem below is by letting the ``complicated 
mathematical structure'' or the worldmachinery represent the {\em whole
time development of the universe} and not just the status at a specific 
moment of time. Then we even open up for effects going backward 
in time. Under such conditions locallity is  no longer fully valid 
and informations about what really happens could sometimes be 
determined from information hidden in the future somehow. You may at least 
see easily that once we in principle open up for laws fixing the future 
so that  it has to be arranged to happen to fullfill some requirements in 
the future, 
then we have ``backward causation'' (i.e. things may happen {\em in order} 
that something having to happen in the furture), and then locality 
makes little sense. A momentary exchange of information faster than 
light could namely then be realized by the information going to the future,
from where it then by ``backward causation'' went back in time but to 
(or at) a different place than from where it came. 

That the quantum mechanics troubles put by Bells 
theorem\cite{Bth} are avoided 
once we have a time perspective, in which all the time development 
is already settled, were already explained by Bell himself on BBC
\cite{Bell} and is known as ``superdeterminism''.
In fact it is in the model\cite{own} presented in the 
present article even 
suggested that we {\em in principle} - but not in practice - except for 
very special cases e.g.involving Higgs particles as we shall see - can 
calculate everything that happens, because 
we have in principle a model for the initial conditions, too.
Rather we should say that the model of Ninomiya's and mine predicts 
in principle, what really happens! Stated in this way it obviously 
means that even the outcome of a measurement - which in usual quantum 
mechanics is supposed to be something only predicted in a statistical way -
gets by us in principle calculable. We must of course seek to deliver some 
arguments for that we {\em in practice} can use statistics, and 
what statisical distribution we predict (we so to speak must 
derive the Born probability distribution from some reasonable arguments)
in much a similar way as to, how one argues for statistics to be applied to 
the amount of poeple comming into a railway train or to the distribution of 
probability for the various numbers of eys on a thrown dice.
In this way we may bring in a kind of {\em unification} of the use of 
statistics 
in quantum mechanics and in other more practical and classical 
applications of statistics. Remember that in other 
philosophies of quantum mechanics the statistics of quantum mechanics 
is a fundamental fortuitousness 
\cite{BohrUlfbeck} contrary to the classical cases of statistics,
which may be a practicel approximation to what could sometimes in principle 
have been calculated.(With the appropriate knowledge of the starting 
conditions of the thrown dice one could calculate the number of eys being 
shown at the end. But a  rather tiny uncertainty makes the distribution 
expected a statistical distribution; by some argument of (practical) ergodicity
one might show that the in this way calculated probability soon would approach 
the usually assumed one, that all the six possibilities for the number of eys 
become of probability just 1/6).

\subsection{The problems}
The real problem for quantum mechanics might be thought about like this:
If one takes the point of view, w.r.t. the concept of time, expressed 
by slogans like ``the future does not exist''; the future rather only develops 
from the 
present, then you may uphold still the Schr{\o}dinger equation
for developping the wave function.  But really inconsistently with the 
Schr{\o}dinger equation do we have in the case of a measurement 
a different development, in which a single measurement result gets 
statistically selceted to be the {\em only one} realized. 
It is  a machinery for delivering the extra information selecting 
which measurement results that shall be provided by the hidden 
variables in hidden variable models.
% come out that is to be provided by the 
%hidden variables in hidden variable models. 

One can say, that it is the point of the present article\cite{own} to 
use a system of degrees of freedom placed  in {\em future} in the 
sense it is determined via calculations involving potential futures,
to settle what result of the measurement shall come out as realized..
This come about like this:
We use in our model  
%in the sense that it is determined from the integrand of the 
%of
 the Feynman-Wenzel-Dirac path way integral - which in our
model has an action with an imaginary part of the action too,
and it is indeed this imaginary part that is important here -.
and then the imaginary part of the action will provide different 
weights to different potential future histories. Then those potential 
futures that
get the highest weight - and that means really the smallest $S_I[path]$,
because the  integrand (that get squared to give the probability)
is $\exp(\frac{i}{\hbar}*(S_R[path] + i S_I[path]))$, will be selected 
as the realized one - will be the ones that should be conceived of as 
the ones realized. This is supposed to mean that essentially only
one measurement result gets selected. We shall return to the estimate 
that indeed the difference in $S_I[path]$ between paths for different 
measurement results will be so big, that one measurement result indeed 
takes practically completely over. This has to do with that a measurement 
enhances the effect so that different results of the measurement 
cause quite different futures.  
(This means, that we have a model with future influencing past,
an idea we have had in connection with baby universe theory \cite{baby}
and the present author  earlier\cite{old}\cite{vacuumbomb} especially w.r.t.
coupling constants getting influenced).

Let us here mention that such influence from the future easily pops 
up in models with time machines for instance caused by worm holes.
An example of worm hole effects popping up even at the LHC machine is
found by the work of Volowich\cite{Vol} in this very proceedings. 

Another problem related to the first mentioned is, that 
as may be seen from the unusual type of logics, that is used 
to describe quantum mechanics, it is not possible to identify 
a quantum state with a state of a system of classical objects 
about which everything can be described by answers to yes-no-questions. 
% information about a quantum system 
%as we possibly can have it, the logic about a quantum system is not 
%as it would be for a sytem were in just one welldefined way.
This means that we cannot describe the state of a quantum system 
by some ``das Ding an sich'' description. We cannot have a model 
as an existing object describable without mixing up with the 
information about the system with our knowledge about it.
Thus if we should indeed attempt to have a  model description by some 
mathematical 
object - ``a mathematical structure '' - to describe  a quantum system,
we are in the trouble, that we cannot do that without including 
into the description some information about our knowledge 
about what we know about the system. 

So wanting, as I want in the project of random dynamics, 
- a major subject in the present article -, to have a description 
of the noumenon (= ``das Ding an sich'') to be a ``mathematical structure''
(being fundamental and without any knowledge about our knowledge about 
it included), we have essentially a no-go w.r.t. to having consistence 
with quantum mechanics! 

The way out suggested in this article is indeed this attempt to a way out:
The problem with the state of a quantum system not being a state that
can be identified with a system being a welldefined mathematical structure,
may be due to that we assume that we shall consider only 
{\em the state at a certain moment of time}. If we, however, think
about rather than to want to describe only the state of the world or 
a system in a special moment of time to describe {\em the whole time
development from the beginning to the end}, i.e. including also the future,
then a priori there might be a chanse of a description by a 
mathematical structure.
That is to say that we should decide to be satisfied to have a 
``das Ding an sich'' model {\em only for the development through 
all times}, but give up to have a ``das Ding an sich''-existence 
of the state at a single  moment of time only. 

Let us immediately reveal, that it is our idea to let the 
Feynman-Wentzel-Dirac pathway integrand - or rather some 
quantities being proportional to the square of this integrand -  be the 
noumenon or ``das Ding an sich''.
In principle we should then be able to see from such a theory
that in practice we should get effectively the usual quantum 
mechanics with its
statistics - basically the Copenhagen interpretation hopefully.        

That there is indeed good hope, that giving up to require the existence 
in a ``das Ding an sich''-sense (i.e.completely objectively) of the world or
the system at a specific moment of time, and only require it for the 
development through all times,may be helpfull to solve the problems 
of the EPS-discussion, 
may be seen by thinking about that the 
time concept is quite crucial in the EPR-discussion. 
In fact there would be no problem in the EPR if we had a theory 
with backward causation fundamentally. It is clear, namely, that 
if indeed we had fundamentally a theory with the future influencing 
the past or arranging the past, then there would be no trouble in
information seemingly being transfered faster than the speed of light.
The information could in principle go forward into the future and then 
back again by the backward causation. May be the best way of seeing 
that this kind of EPR-related problems disappears in the case of 
a model with backward causation is by refering to the Bell-type 
superdeterminism\cite{Bell}. Indeed if we have a model like 
Ninomiya's and mine 
with  the complex action, in which all what happens is in principle 
calculable, then the argumentation in EPR or say in Bells theorem\cite{Bth} 
of one being able to choose to measure different things (depending 
on some ``free will'', one could almost say) disappears. 
This is simply because the experimentalists cannot do anything else than, 
what they could in principle have been calculated in our model to
have to do. Everything is in principle calculable, both 
what the experimentalists will choose to measure -on both of 
particles which were prepared in the entangled way -  and what will
be the result of those measurements. Thus the discussion of the 
contrafactual possibilities for measurements are out of question;
you could just have calculated what would be measured and what is not 
measured  in reality, but only in a contrafactual history,
is not relevant.

%\subsection{Direction of our solution, use future as hidden variables}

\section{Random dynamics, first very speculative start}
Let us at this point give a quick review of the project of Random Dynamics, 
which is dreamt to be a theory of everything (T.O.E.). It is based on the 
speculation
that we can assume almost whatever theory it should be, if it sufficiently
complicated, and then we should be able to show that the effective model 
resulting will almost not depend on any details. Actually it is 
hoped that almost any complicated theory would  turn out to be 
the physics, we know from experiment. You could put it as the slogan:
{\em Almost all theories would turn out right with respect to the effective 
laws predicted.}

To at least give an idea about the start of this project (random dynamics)
let us mention, that I usually call the start-assumption ``a random 
very complicated mathematical structure''. To make this concept 
just a bit concrete let us describe it by the following 
computer-set-up:

Imagine we have a relatively simple computer (but it hopefully 
does not matter how it precisely is) and give it a very long 
random input string (a random tape),from  which it then produce an 
output tape,which is supposed to be exceedingly much longer 
than the input one. Both the input and the output strings are supposed 
to be enourmously long (enourmously many bits), but we assume that the 
output-string  to be really enourmously much longer 
than the input-string, not only by a moderate factor, but rather so 
that likely the output-string may have a length, that could be like an 
exponential of the length of the input-string. Then it is clear that there
must be an enourmous amount of regularities in the output-string.
Even though the input-string is completely random, just noice, then 
this amount of noice is compared to the information on the output-string
extremely small. Thus most information on the output-string must be 
regularities, a kind of repetition of the same again and again in different 
variation of the theme so to speak. It is now our point with this 
computer-analogy to identify this output-string with the ``random
mathematical structure'' which in turn is identified with the world
(or as we already alluded to strictly speaking rather the whole 
{\em development of the world}).

According to the discussion already alluded to we shall indeed 
let the output-string describe not the world at a given moment but 
rather the development of the universe all through times. In this sense the 
output-string should represent all what happens at all times in some way which 
of course could be not totally trivial.

\subsection{Second step in Random Dynamics}
The hope of Random Dynamics now is, that we by arguing little by little 
can find something on such an a priori difficult to read output-string 
to identify with features of the known physical world. The success of
random dynamics would mean, that we could propose such a series of 
identifications of features of the output-string structure with features 
in the physical world, that we would get rules among the physical 
features from the rules deduced from the corresponding features in the 
output-string, and then the hope is, that  it would turn out, that these 
rules were indeed 
identifyable with (some of) the laws of nature already known. That is to
say, we hope that with the appropriate identifications of physical concepts 
with features of the output-string we can deduce the various branches and 
laws of physics.In fact we hope that we shall find a series of deductions 
of the variuous elements in the physical theory we know, of laws let us say,
so that we at the end we would  have deduced essentially all and stand say 
with the Standard Model as a derived result.                  

In the second step\cite{Bled} - just after having assumed the huge 
output-string 
being the fundamental structure - we search to realize the hope of 
the starting point not being important (remember: we hope in Random Dynamics
to get the same resulting effective emerging theory almost independent 
of the starting model!) by finding a way of formulating the output-string 
into a language, that could be used on ``practically everything''. We seek to 
realize this formulation in terms of a notation, that can be used on 
(almost)``everything'' by starting by using some of the most fundamental 
mathematical 
concepts - the numbers - for the general description of the output-string.

Actually we can even very easily argue, that having to make a description 
of the enourmously big output-string - so as to bring at least some order 
in it to be recognizable -  the first thought would be in such an as
already mentioned very repetitive structure to characterize it by 
talking about and defining some ``substructures'' - i.e. some 
large combinations or patterns defined some way or an other, so that 
they do not have to be localized to special regions on the output-tape -
and then count those so as to describe the whole output-string approximately
by delivering the numbers of each of the many different types of 
``substructures'' found.  The reader should notice that just by 
making this choice of using a description in terms of numbers of 
``substructures'' one has achieved to represent the information about 
the output-string, that is being  kept, by a point in 
coordinate system. In fact the   coordinates of this point  are 
in correspondance 
   with the various recognizable ``substructures'', and the coordinate value 
should essentially describe the number of substructures of the type in 
question.
 At first of course 
the numbers of the various ``substructures'' must be non-negative integers.
Then, however, we shall successively almost follow the introduction 
of the variation types of numbers: natural numbers, integers, real numbers,
complex numbers. That is to say we come with arguments, that although we
started by the description of just non-negative integer numbers of 
``substructures'', then we can develop the description by slightly 
changing it so as to get successively the integers proper etc in.
The first step in this argumentation is e.g. that one argues: 
Let us go to a  description wherein we only consider the deviations of 
the numbers
of ``substructures'' relative to some``normal number of them''. That is to say 
we imagine some ``relativly simple'' ``background state of the sytem''
in which the number of ``substructures'' have some background values.
Then we redefine to consider intead of the absolute number of a 
``substructure'' as the coordinate the difference of this number relative 
to the number of the same ``substructure'' in the ``back ground'' situation.
In this way we can get the new coordinates to be allowed to be both positive 
and negative. So in this way we introduce the full set of integers rather than 
the only non-negative integers, from where we started. 

The step from integers to real numbers may be just argued by saying, that 
of the interesting types - namely the dominant ones - of ``substructures''
there are so many, that we would naturally take a unit which corresponds 
to a very large number of them, and thus the number of units would in 
practice be a real number.

The transition from real to complex we shall not truly make except 
in one case: 

We have to have in mind that these ``substructures'' 
are themselves enourmously complicated and we have to wonder how the 
numbers of a species of ``substructure'' will vary with variations in 
of for what ``substructure'' we want the number. For the copiuos types of 
``substructure'' we might imagine that just that type of ``substructure''
got copious, because it were able - in analogy to biological survival 
of the fittest - to make many replica of itself. We might think of 
some rather similar, but still a bit different ``substructures'' forming
a kind of family with the first one and that they  
would collaborate  with each other in making those 
types be replicated and replicated again and again - we can say under 
the running of the random program in the computer model above.
To be correct we should think of some prestates, in the intermediate memory 
of the computer,  for the structures on the 
output tape being what is replicating itself in collaboration with 
the ``family members'' of related prestates. 
If it now is the case, that such a replication plays a dominant 
role in getting the numbers of ``substructures'', which are copious and thus 
interesting, then going from considering one type of ``substructure''
to an only slightly different type the change in ability of replication 
might change slightly. But since they replicate to an enourmous degree 
even a small change in ability to replicate will cause a change in copiuosness,
i.e. in the number of substructures on the final output-string, 
in an {\em exponential way}. By this we mean that the number 
of ``substructures'' is changed by the same factor roughly for a 
given change in the
replication-ability. If it makes - as we may assume- sense to 
talk about two similar modifications of two different types of 
``substructures'' the modifications of the numbers of these two 
types would be changed by about the same {\em factors}.  This means that
we would think that the {\em logarithm} of the number of a given kind of 
``substructure'' occuring in the output-string would be a more 
smoothly varying function as function of what can be varied in the 
definition of the class of ``substructure'' counted, than 
the number straight away. 
   
The just given argumentation would be right, if we just had one 
type of ``substructure'' replicating alone, but if we take into account 
that they{\em collaborate} with neighbors in the space of types of
``substructures'', then the number of a `substructure'' occuring at the end
could get a more oscillating behavior in addition to its {``\em exponential''}
form. One could imagine that a couple of family members under the 
calculation of the computer were replaced back and forth by each other
- $A$ made $B$ and $B$ made $A$ - so that the number of $A$'s say
would come to oscillate, more like a sine or cosine than like a 
exponential. In addition of course it might grow 
up exponentially. When finally the number of the ``substructures'' 
come out on the final output-string, the number would best be described 
- in order to be smooth and nice - as a functon which is of the character
of sine or cosine shape function multiplied by an exponential.
But that is precisely how one get by using (the real part of) a 
{\em complex number } 
exponetial. So we have here argued, that 
{\em the number of a specific type of ``substructure'' should be 
most smoothly given as} (the real part of ){\em a complex exponential 
of the features of the
structure in question described by numbers}.

\subsection{Series of affine spaces, basis in one, points in the next}     
In the foregoing subsection we argued for that a natural way to 
describe our ``mathematical structure'' (actually taken there to be 
the output-string) were to describe it by counting and to give the 
numbers of a large number of ``substructures''; then by comparing to 
a standard or background one might in stead count the difference in these 
numbers compared to this background, but that were not so important again.
But once we may have convinced ourselves is this proceedure, we might 
immediately perform the same proceedure in describing one of these 
``substructures''. That is to say we should describe the ``substructures'' 
by giving how many ``subsubstructure''s of a given type there are in the 
``substructure'' in question. In this way we would also after having 
introduced a ``back ground'' and a ``unit'' end up with 
having a type of ``substructure'' described by a point in a vector 
space or perhaps better an affine space, in which the coordinates 
count - relative to the back-ground-``subsubstructure''s in the unit 
(of certain large numbers the number of ``subsubstructure''s) in the
``substructure'' in question. We make indeed in this way every 
``substructure''-type being described by the numbers of its different 
``subsubstructure''s. 

Let us summarize that we first proposed to extract as the main features to 
be described in the ``complicated (random) mathematical structure''
identified with the output-string the numbers of ``substructure''s 
of different kinds. Secondly we then describe these ``substructure''s
by the analogous technology, by describing them by means of the 
number of ``subsubstructure''s of various kinds inside the ``substructure''
to be described. The ``substructure''s thus becomes identified with 
the basis vectors in the affine space or vector space in which the 
``complicated (random) mathematical structure'' corresponds to a 
point (namely to the set of coordniates, each of which describe 
the number of the ``substructure''s corresponding to that coordinate or
basisvector). But it(=the ``substructure'') is also on another 
- but analogous - affine space 
(or vector space) a point; it is namely a point on the affine space 
in which the coordinates or basis vectors correspond, each of them, to 
a ``subsubstructure''. We thus have now in our description two
affine spaces, and the basis vectors in the first have come into 
correspondance with the points on the next. 

The reader can imagine, how one can make an anlogous description 
now of the ``subsubstructure''s in terms of numbers of ``subsubsubstructure''s
and so on.

In this way we have argued for that one gets almost any complicated enough 
mathematical structure  - or we could say theory -  may get described 
by a {\em series
of affine spaces} or vector spaces {\em connected one to the next 
by the coordinate basis for one affine space being in correspondance 
to the points on the next affine space in the series.}

It is in the Random Dynamics project now made an identification of these 
successive affine spaces with similar spaces in the phenomenologically 
known physics picture. 

Crudely the identification should run like this:

The types of ``substructure''s in the affine space in which the 
full output-string is reprented by a point should be identified with 
the Feynma-Wentzel-Dirac-paths from the path way integral formulation 
of quantum mechanics. Then the coordinates of the point corresponding 
to the output-string (or equaivalently the ``complicated (random)
mathematical structure'') are to be identified with the 
Feynman-Wentzel-Dirac-pathway integrand
\begin{equation}
``coordinate''(\hbox{related to numbers of ``substructure'' path}) 
= \exp( \frac{i}{\hbar}* S[path])    \label{integrand}
\end{equation}
where we have of course written as usual formally at least 
the integrand in the functional integral as the exponential 
of $\frac{i}{\hbar}$ times the action $S[path]$.

The reader can probably easily by himself see that, if the ``substructure''s
correspond to (get identified with) paths taken in a field theory
as thinkable developments of the fields through all times and space,
then the coordinates by which such a path is described must be 
in correspondance with a cross product of the usual say Minkowski space
(or its general relativity analogue) with some discrete set, corresponding 
to specifying the type and components of the field. In fact the point is
that a path being a field development through time
 is represented by a number- the field value - for every combination 
of a space time point (a Minkowski space point) and an index-value telling
which type of field and which component  is concerned. Thus the path is indeed 
given by a number for each such combination of an index and a Minkowski 
space point (= event). Apart from the detail with the component and 
type of field specifying index it is thus the affine space the points 
of which are the ``subsubstructure''types that is identified with 
our ordinary space-time (manifold). 

You may note that we here have included time on an equal footing with space
and that at this in the Random Dynamics fundamental way of looking at it
all times are already there. There is here no talk about the future not 
existing (yet) but rather that we first a bit late begin to find 
some parameter to identify with time. Following the series of affine spaces 
suggested the time would be a coordinate in the space(-time) of the 
``subsubsturcure''-types. But that would then mean that the time $t$
as a number would correspond to and be proportional roughly to 
the number of (relative to some background) the number of a specific type 
of ``subsubsubstructure''s in the ``subsubstructure'' corresponding 
to the event in question. There is no time in the deep fundamental level
of ``das Ding an sich'' before we get a certain ``subsubsubstructure''
identified as being the special one that by its number of occurences 
make up the time $t$. In really devlopping the model we should 
presumably say that this number of the specific type of 
``subsubsubstructure''s
is rather a sort of pretime because it could get modified by 
some dynamical metric tensor like in gravity which we hope to get 
out almost unavoidably although we may admit that we should not claim that
we really succeeded in getting it out convincingly yet.

But we have at least reached a suggetsive model in which the time 
concept does not come in immediately at the most fundamental level.
Thinking close to our fundamental picture one should best think
in a timeless way: all times exist and thinking from the almost God-way
of our fundamental, all times are real. Then at some level one identifies 
a prototime as the number of a certain special ``subsubsubstructure'',
and  then we must imagine that the conditions end up being so that 
we as being present in terms of ``subsubstructures'' with this 
prototime number having a given value only have good information 
about ``subsubstructures'' with say a lower number of this specific 
``subsubsubstructure'' representing the time. Then we might 
``at that time'' easily say that the future meaning ``subsubstructures''
with a bigger number of the specific ``subsubsubstructure''
representing time concept is so unnknown that we would consider 
it ``not yet existing'' psycologically.
%   way  they in human language 
%will turn out or have turned out some day.

Already at this way of thinking you might see that the Bell-superdeterminism
\cite{Bell} looks not so far out:
How the whole complicated mathematical structure comes to be is 
a result of the calculation in our computer model with the random 
input-string, and there is at this fundamental level no 
time and no restriction for how things develop in time at first.
We must hope and work on a way to derive ``in practice'' such 
restrictions that the whole things come to look as a system developping in 
time from a relatively simple start. We must hope that we can derive 
that poeple living in such a world will get a time feeling and 
might even come to believes due to their practical experience that 
the future hardly exist in some way which one can call it does not exist 
yet. But all this is to be derived later in the logical development;
if one cannot derive it, it would mean that the hoped for Random Dyanmics
project failed. But we have not given up yet, but rather claim it 
to look promissing that we shall derive it.  
 
\section{The lack of reason for the action being real}
Even with a great amount of optimism concerning the above dream of 
deriving as almost unavoidable a complicated mathematical structure 
being identifyable with a Feynman-Wentzel-Dirac path way formulation 
of a quantum field theory there were one thing, we did not derive at first:
we got the action {\em complex}, but in usual theory the action is real.

Actually in the argumentation above for having to excuse making 
complex numbers come in in  order to describe as exponenatials the oscillatory
sine or cosine like behavior meant that the intergand (\ref{integrand})
were argued to be complex rather than real; but that means that 
what we had to argue especially for were that the action $S[path]$ 
became complex rather than just {\em purely imaginary}! So even if this 
 special argument for finding that the action would be complex rather than 
just imaginary were wrong we would not get the action real as needed 
for phenomenological success but rather get it purely imaginary instead.  

Phenomenologically having a purely imaginary action would 
of course be without any hope of working, while having a complex action 
which at least has the observed real part has a better chance than that 
at least. Indeed we shall argue that it is quite imaginable that 
if the acion in nature had indeed both a real and an imaginary 
part then it could go so that in practice we would almost 
only ``see'' the real part, and the imaginary part would rather 
fix the inital conditions, but even that in a way is  not necessarily 
in disagreement with what we see.

\section{Essential fixing of the whole history of the Universe}
At first it looks that having an imaginary part of the action in say the 
Standard Model might drastically change even the equations of motion,
and that might be catastrophic for our model phenomenologically. 
However, we think that we can at the end get the effects of the 
imaginary part be talked down to be small, even if they at the first 
are as big as expected from the phase of the coefficients on the various
terms being roughly random. The argument may not be quite watertight and we 
still work on it. It contains the following ingredients:

1) If the imaginary action were small of first order then 
the deviation of the effective classical equation of motion 
from the one obtained alone from the real part of the action 
alone would be only of second order.

2)In addition to the modification of the equation of motion 
(which is expected at the end to be small) the imaginary part of the 
action has the effect of determining the (say the classical)
solution among the possible ones obeying the equations of motion 
to be realized. In fact the result is approximately that the solution 
being realized is that one which has the smallest(rather most negative)
value of the imaginary part $S_I(solution)$.

3)Such a determination of the realized solution - meaning that 
the initial conditions gets determined from our model - in a way 
that even depends on what in the solution goes on at all times, 
both past and future,means that  the model tends to make arranged 
events happen, if they somehow can produce a big negative 
contribution to the imaginary part of action $S_I$. In this 
sense our model tends to have such arranged events, which would look 
like miracles (or anti miracles, if they are bad events).

4) But experimentally miracles and anti miracles are very seldom 
- if they occur at all - and now we have an argumentaion to explain them
to be indeed seldom: Because of the era of times in which we can truly 
recognize when miracles or anti miracles occur is very small compared
to the total time over which the universe exist very many miracles 
are likely to occur at times different from the times in which we would 
be able to know about it and recognize it as miracle or anti miracle. 
We 
%Thus we 
must namely  
understand that because of the need for the equations of motion being satisfied
(determinism) the initial conditions cannot so easily be adjusted 
to make miracles or anti miracles in all the many rather small 
eras in which we may detect them. Thus we must admit that the number 
of arranged event that can be detected in practice will be depressed 
the longer the total life time of the universe is compared to the human
scale. Further this possibility for getting observable 
arranged events (miracles or anti mircles) is reduced by the 
likely speculation that in some era close to the big bang era,
or in the inflation era say, the contribution to the imaginary part of the 
action has a very high order of magnitude. If so then namely such an era 
could contribute so strongly to the imaginary action that it becomes 
what happens in this era that determines almost the whole inital 
condition. This of course would be phenomenolgically the best scenario 
since what we see phenomenologically is that only the very early time,
``the beginning'', is the era in which the intial state is 
fixed in a simple way. When I here have put the word ``the beginning''
in quotation marks, it is because a bouncing Universe starting at time 
being minus infinity with a hugely big universe which then {\em deminizes}
become relatively very small and then grows again towards our time and then 
even further, is in our model more attracktive than in other models.
While we have the second law of thermodynamics as a derived law that only may 
work during the expanding era, a theory with the second law of thermodynamics 
as a fundamental assumption to be valid even in the universe decreasing 
era would make it very difficult to have a working scenario.

The question whether we can indeed get our model make the phenomenologically 
attracktive result that the era, when the Universe were small - i.e. 
crudely around big bang time (if there really were a singularity 
of course a little after big bang) - to dominate the selection of the 
solution to be realized depends on whether this era around big bang time 
gives the order of magnitudewise  biggest contribution to the imaginary part 
of the action $S_I[history]$. It is namely suggestively argued that 
what determines the possible history that shall be realized is this 
imaginary part, in fact the realized solution is seen to be the one 
that causes the imaginary part to be minimized. One may see that 
it is such a minimization of the imaginary part that gets realized 
without even going into details in how         
%It is really not difficult to see the main point even without going in detail 
%with how 
we shall interprete our model with complex action.
It is enough  if we just think as we 
suggest to interprete our model with complex action by means of the 
Feynman-Wentzel-Dirac-path way integral formulation. In fact 
we only have to have in mind that the integrand in this path-integral 
$\exp(\frac{i}{\hbar}*S[path]) \propto \exp(-\frac{1}{\hbar} S_I[path])$
goes exponentially down the more positive the imaginary part $S_I[path]$.
This would, almost whatever the interpretation might be, mean that 
unless the imaginary action is so small (so negative) as possible, 
the contribution to the Feynman-Wentzel-Dirac path way integral will 
be very small. Thus only the smallest possible $S_I$ achievable will  without
other reasons for severe suppression  be the dominant contribution.
Thus we see, that what can come to be important, and that must correspond to 
what we shall conceive as being realized, must be the contribution 
- among the say classical solutions possible - with the smallest imaginary 
part of the action.

One shall imagine that it is at the end a rather strong selection 
of almost only one classical solution to be the realized one.
That is a very welcome prediction from the point of view 
of our usual experience that essentially only one history of the 
universe gets realized. Here of course we must admit that although 
we all essetially feel that only one thing happens, then we know that somewhat
seldomly though there are the double slit experiments. In some experiments 
it is not possible to say through which slit say a certain particle has passed.
Even in hind sight - i.e. after the experiment has been completely finished -
we cannot say through which of the slits the particle went. 

So we may form ourselves the following picture of our model 
prediction:

Due to the minimalization of the imaginary part of the action an almost 
unique classical solution is selected as the realized one. Estimates of the 
likely distance in $S_I$-values for the various possible calssical solutions 
that do not follow each other along long time interval is that they are so 
big,  that the next to lowest $S_I$ cannot compete at all with the 
lowest one. Only in the case that a couple of solutions follow each 
other through almost all times shall we expect that the imaginary parts 
of the action could be so close that both partners in such a pair 
could be significant (together).  

\subsection{But we can still have some almost classical paths interfere}
When we have such mostly each other following classical paths 
contributing bunches of paths, then it will function as 
interfering paths and can deliver the interference in such experiments as
e.g. the double split experiment.
In this way our model should achieve to make almost a classical 
history being selected uniquely, but with a somewhat seldom possibility 
of this classical path being split a bit for shorter periodes. 

\section{The reason why a measurement causes no more interference}
Now of course we want to see what happens in a measurement:
Let us first note that a typical measurement consists in an enhancement.
In some way or another a small effect - a single particle or so - 
is arranged by the constructor of the measurement instrument to
be enhanced and make a big effect. Typically some system is set up, 
which is in some metastable state, which then by being touched rather weakly
can go into a different state, that deviates a lot, so that the effect 
of the quantity being measured gets a very big effect (later).
It is of course also so that depending on the result of the 
measured quantity the cascade will go differently, and so the difference
between the state of the universe after different measured results will 
be very different. 

This now means that the development reached after different 
measurement result are quite different, and thus the future contributuions
to the imaginary action will depend on the measurement result.
It does not depend on the result necessarily in a systematic way,
but we shall rather think of it that each possible result 
value (= eigenvalue of the measured operator) gives an in practice big 
random number for the future contribution to the imaginary part of the action.
In this way it will almost certainly be {\em only one} of the eigenvalues, 
that will in practice contribute to the total Feynman-Wentzel-Dirac 
path way integral. That means that in the philosophy of ``what really
happens'' being determined by some second order effect of the 
path way integral integrand, all that gets realized will be what 
corresponds to one single measurement result.
The just given argument is suppoed to be the argument in our complex action 
model for the from a pure Schr{\o}dinger equation point of view rather 
mysterious phenomenological fact that only one measurement result gets 
realized. In this way we want to say that our selction of an 
or a few surroundings of classical solution tracks (system of paths)
by the effect of the imaginary part of the action in fact leads to
- from the argument of the strong dependence on the measured result
due to the enhancement - that only one result gets realized. 
This fact of the Copenhagen interpretation is thus here claimed to 
come out as an approximate result (actually in an extremely good 
approximation). 

If however, two solutions almost follow each other through almost 
all times and only deviate in a very tiny time interval, then 
there is a better possibility that they could have indeed very close 
or practically the same imaginary parts of the action. So in such 
cases the appearance of contribution to the functional integral 
from both solutions is not totally excluded. In this sense interference is 
possible in our model. It could be that two classical solutions
could come so close, that they have a chance to give competitive 
contribution and thus to make interference occur. 

\subsection{Quantum mechanics support of our model}
We would like to come with the remark, that indeed something much 
like our type of model is almost called for, if one does not accept 
the many world interpretation  and want to make a picture 
with reductionism be consistent. The point is to think 
about that the property of certain process being indeed a measurement 
will not necessarily be obvious from the start. A measurement so to speak takes
time. Nevertheless we should also imagine that the measured particle so to 
speak has the measured eigenvalue as its position or momentum or whaever 
were measured from the time, that the measurement has begun. Otherwise 
we have to live with the many world intepretation and a superposition 
of two parts of the measurement apparatus being excited in an entangled way
at least as long time as the important part of the measurement 
takes place.

If one want to avoid such problems one has to allow for 
the effect of whether we have indeed a measureent or not 
to propagate backward in time!
To have to have such a knowledge about whether it ends with a measurement 
preferably propagate backward in time really is a strong call 
for a model like ours in which there are backward causation effects. 
We simply claim to need such effects in the case of slowly going 
measurements if we shall avoid the many-world-scenario.

We want to claim that the fact that we get in our model an easier
picture for the measurement problems should be taken as a point 
of evidence for our model.

\section{Conclusion}         
We have reviewed some initial steps in the ambitious project of Random Dynamics
leading at least to the possibility of the hope for getting out naturally
from almost any starting ``complicated (random) mathematical structure 
with a lot of regularites though'' a Feynman-Wentzel-Dirac- path-way formulated
quantum mechanics model. However, and that is a very important part
of our work: We fail at first at one point: although we have some argument 
for how to get the functional integrand $\exp(\frac{i}{\hbar}*S[path])$ 
become complex, we lack an argument for {\em why the action $S[path]$
should be real} at first. 

This lack of success in the Random Dynamics project we take as 
the suggestion to study what would happen, if the action were complex.

It turns out, that we can essentially get the effect of the imaginary 
part get both suppressed and largely restricted to make this imaginary 
determine the initial conditions rather than the equations of motion.

But this acceptance of that there {\em is} an imaginary part 
of the action being different from zero, although we have mechanisms 
that can suppress it, has great significances:

1) Our model becomes ``superdeterministic'' in the sense that 
everything that happens get in principle calculable, not only after 
having got the information about the beginning, but rather {\em calculable
from scratch provided we just know the various complex coefficients
in the (complex) Lagrangian density}. So everything is determined
``in advance''(our picture is in the first levels ontill a time concepts 
is found, timeless, and it is best to imagine in our model that ``God''
thinks in a timeless way, so that both future and past and present
exist).

It should be admitted though, that there is allowance for a little 
loophole, if one thinks classically in this total preditivity of 
everything:

There is an opening for paths not representing different maesurement 
results of any measurement to be seperated for some (not too long)
time.In fact for instance the famous double slit experiment 
in which we never shall know through,through  which of the two slits 
the particle went (if it were not measured) we shall even in our model 
even in principle not be able to calculate through which slit it went.
\footnote{We might though define what we could call the 
hindsight point of view: Instead of asking for useful information 
you ask for the metaphysical one and accept that just when the 
position of a particle were measured, I could still pretend 
in words, that it would have the momentum with which it were prepared.
However, even in such a hindsight point of view sometimes the 
slit, through which the particle past, will not be definite.}

2) Especially it is in principle calculable which detector 
will click, i.e. what will be the outcome of a quantum measurement.
On the contrary in the  
usual theory it is supposed to be given completlely randomly,by 
statistics only.

3) But this also means, that it is explainable due to the enhancement
connected with having a measurement instrument for elementary 
particles, atoms etc. that (with very high accuracy) we  only get one 
measurement result. This result of only one measurement is the
only possibility that  will occur with any significant contribution 
in the Feynman-Wentzel-Dirac-path way integral (from which 
the variables counting as noumenon depend as second order expression).
Thus our model can be claimed to explain the fact that only one
measurement result occurs, a fact often considered rightly
quite mysterious.

It is our main point, that the from the usual theory and time concept point 
of view strange model of ours is actually advantageous w.r.t. consistency 
with quantum mechanics. So we would claim that the problems of quantum
mechanics are actually indicators in the direction of precisely our 
model. 

Only if the contributions to the imaginary part of the action get exceptionally
large, do we expect that they will give so strong effects that we 
would e.g. see that to some extend the initial conditions have been 
finetuned so as to arrange for events with exceptionally high negative 
imaginary action or prevent events with exceptionallly high positive 
imaginary action. Neverheless we believe to have spotted cases of 
such prearrangement in connection with production of Higgs particles,
and we also think that one might very easily put up phenomenological 
couplings for the inflaton-field say such as to make the inflation periode
 become a case of prearranged very negative imaginary action comming about.
Indeed our speculation is that due to that the imaginary part of the 
coefficient $m_{Higgs}^2$ to the numerical square $|\phi(x)|^2$
of the Higgs-field $\phi(x)$ -  let us call it $m_{Higgs}^2|_I$ - 
is much bigger than the corresponding real part $m_{Higgs}^2|_R$
(which is suppressed we think due to some mechanism solving the 
``scale problem'', closely related to the hierarchy problem).
We here have put $m_{Higgs}^2 = m_{Higgs}^2|_R + i m_{Higgs}^2|_I$. 
And then we easily get that the supposedly effective huge term
$i* m_{Higgs}^2|_I |\phi(x)|^2$ in the Lagrangian density
only in practice shows up, when we have genuinely produced 
Higgs particles, in an accelerator say. Because it seems that 
genuine flying around Higgs particles are not favoured 
to be present in Nature - we did not see any so far - 
presumably the sign of the term, that is huge is so, that 
Higgs particle production get disfavoured on the realized history 
of the universe. Thus our prediction becomes, that accelerators, 
that produce large amounts of Higgs particles, should have bad 
luck and for instance be prevented from working by the US-Congress,
as it happened in the case of the SSC(=Superconducting Super Collider.).
About 21 km of tunnel is still there, so it were a rather remarkable case
of closing an expencive machine at that stage, it were almost an anti-miracle.
Also LHC(=Large Hadron Colliderat CERN) had bad luck just in the 
month it were about to start up, Oktober 2008.    
%{\bf The references below are at first only the references from a 
%random article on a similar topic and shall be modified a lot:}
An explosion has now delayed the start by at least a year.

%\thispagestyle{empty}
%\section{Our monopole coupling versus $d$ curve and successful agreement}

%\begin{tabular}[h]{|l|c|c|c|c|}
%\hline
%Group &$ d_0$ &$ d$ &$ 3\tilde{g}^2/\pi|_{pred}$ & 
%$3\tilde{g}^2/\pi`_{exp}$ \\
%\hline
%$U(1)$ & 0 & 0 & ...& 27.7 \\
%$SU(2)$ & $11/3 =  3.67$ &$3.67$& ... & $196._{12}$\\
%$SU(3)$ & $11\sqrt{65/108}=8.534$ &...& ...& $212$\\
%\hline 
%\end{tabular}
\end{document}